\documentclass[compsoc, conference, a4paper, 10pt, times]{IEEEtran}

\usepackage{float}
\usepackage[hyphens]{url}
\usepackage{graphicx}
\usepackage{mdframed}
\usepackage{multirow}
\usepackage{microtype}
\usepackage[table,dvipsnames]{xcolor}

\usepackage{tikz}
\usetikzlibrary{decorations.pathreplacing}
\usetikzlibrary{positioning}
\usetikzlibrary{patterns,shapes,arrows}
\usetikzlibrary{fit,calc}
\pgfdeclarelayer{background}
\pgfdeclarelayer{foreground}
\pgfsetlayers{background,main,foreground}

\usepackage{listings}
\lstdefinelanguage{Cmorekeywords}{
  language = C,
  basicstyle={\footnotesize\ttfamily},
  numbers=none,
  numberstyle=\tiny\color{gray},
  keywordstyle=\color{blue},
  commentstyle=\color{red},
  stringstyle=\color{OliveGreen},
  morekeywords = {
    uint8\_t,
    uint16\_t,
    time_t,
    std,
    string,
    PRInt32,
    regex,
    smatch
  }
}
\lstnewenvironment{Ccode}
                  {\lstset{language = Cmorekeywords}}
                  {}

\usepackage{listings}
\lstdefinelanguage{Lmorekeywords}{
  language = {[5.0]Lua},
  basicstyle={\small\ttfamily},
  numbers=none,
  numberstyle=\tiny\color{gray},
  keywordstyle=\color{blue},
  commentstyle=\color{red},
  stringstyle=\color{OliveGreen},
}
\lstnewenvironment{Lcode}
                  {\lstset{language = Lmorekeywords}}
                  {}

\begin{document}

\title{Assessing and Exploiting Domain Name Misinformation}

\author{\IEEEauthorblockN{Blake Anderson}
\IEEEauthorblockA{Cisco\\blake.anderson@cisco.com}
\and
\IEEEauthorblockN{David McGrew}
\IEEEauthorblockA{Cisco\\mcgrew@cisco.com}}

\maketitle

\begin{abstract}
Cloud providers' support for network evasion techniques that misrepresent the server's domain name is more prevalent than previously believed, which has serious implications for security and privacy due to the reliance on domain names in common security architectures. Domain fronting is one such evasive technique used by privacy enhancing technologies and malware to hide the domains they visit, and it uses shared hosting and HTTPS to present a benign domain to observers while signaling the target domain in the encrypted HTTP request. In this paper, we construct an ontology of domain name misinformation and detail a novel measurement methodology to identify support among cloud infrastructure providers. Despite several of the largest cloud providers having publicly stated that they no longer support domain fronting, our findings demonstrate a more complex environment with many exceptions.

We also present a novel and straightforward attack that allows an adversary to man-in-the-middle all the victim's encrypted traffic bound to a content delivery network that supports domain fronting, breaking the authenticity, confidentiality, and integrity guarantees expected by the victim when using HTTPS. By using dynamic linker hijacking to rewrite the HTTP Host field, our attack does not generate any artifacts that are visible to the victim or passive network monitoring solutions, and the attacker does not need a separate channel to exfiltrate data or perform command-and-control, which can be achieved by rewriting HTTP headers.
\end{abstract}


\section{Introduction}

Domain name-based intelligence has long been used by the security research community to identify and remediate malware infections and other attacks. For example, Ma et al. extracted domain names from Mirai binaries and then used passive and active DNS datasets to perform DNS expansion in order to construct a graph highlighting the shared infrastructure used by Mirai variants \cite{antonakakis2017understanding}. Roberts et al. use domain names in TLS certificates to identify domain impersonation attacks \cite{roberts2019you}.

The above investigations relied on the fact that the domain names in DNS responses and TLS certificates accurately represented the identity of the servers that the clients intended to communicate with. This assumption is not necessarily true. Evasive software can misrepresent its servers' domain names in DNS and TLS, a feature utilized by both privacy enhancing tools and malware.

Domain fronting is a popular method that applications use to misrepresent their target server's domain name. It leverages shared hosting and HTTPS to present a benign domain in the DNS request, TLS \texttt{client\_hello}, and TLS \texttt{certificate}, while signaling the target domain in the encrypted HTTP request. Domain fronting is possible when shared hosting providers use a TLS termination proxy, providing themselves visibility into the HTTP \texttt{Host}. Section \ref{sec:ontology} provides an in-depth description of domain fronting and related techniques.


Many cloud providers have stated that they no longer support domain fronting \cite{google2018df,amazon2018df}. To verify these statements and to better understand the domain name misinformation ecosystem, we developed a novel measurement system that identifies candidate sets of domain name/IP address tuples related to each other by increasingly specific measures. In this paper, we first analyze candidate sets based on their autonomous systems. We then use the insight that, when a provider supports domain fronting, the target and front domains are both from a specific set of domains associated with that provider. These sets of candidate domains can be constructed through passive DNS monitoring, and then scanned to characterize what domain misinformation techniques, if any, are supported. We construct DNS-related candidate sets using the domain name and fully qualified domain name (FQDN) returned in DNS CNAME records.

As an example, the measurement system identifies candidate sets related by Fastly using the \texttt{FASTLY} autonomous system, the \texttt{*.fastly.net.} canonical domain name, and the \texttt{j.sni.global.fastly.net.} canonical FQDN. As we demonstrate in Section \ref{sec:measurement-results}, the increasing level of specificity allows one to have a much clearer view into the conditions in which hosting providers support domain name misinformation.

Once we have the candidate sets of domain name/IP address tuples, we then investigate pairs of tuples belonging to the same set to identify support for domain name misinformation. For each pair of destinations, several scans are initiated to retrieve baseline values as well as to exercise different techniques such as domain fronting. We present the scanning methodology in Section \ref{sec:measurement-methodology}.

Our results show that many cloud providers support domain fronting; sometimes intentionally, sometimes optionally, and sometimes unwittingly. For example, while you cannot use \texttt{www.google.com} to front domains hosted by Google App Engine, the set of domains mapping to the \texttt{ghs.googlehosted.com} canonical name creates an equivalence class of domains that can all be used to front between each other.

Complicating the situation further, many popular services are hosted by multiple, unrelated providers. For example, \texttt{ctldl.windowsupdate.com} is hosted by at least Azure, Akamai, StackPath, and Limelight. While you cannot perform domain fronting with this domain through Azure or Akamai, you can through StackPath and Limelight. A security model that trusts the domain name, without considering the provider and its support for domain evasion, is vulnerable to evasion.

After measuring support for domain name misinformation, Section \ref{sec:attacking-df} shifts the focus towards how an attacker can co-opt domain fronting, which was put forward as a privacy enhancing technology that allows political dissidents of repressive regimes access to an uncensored Internet \cite{fifield2015blocking}. While these privacy goals of domain fronting are unequivocally altruistic, it would be irresponsible to ignore how these same regimes can leverage domain fronting to stealthily maintain a surveillance state.

A feature of modern content delivery networks (CDNs) is the decoupling of a domain's TLS certificate and origin server. When the CDN does not verify that the domain name appearing in the HTTP \texttt{Host} field is represented in the certificate, the core tenet of trust between the user and origin server is broken. In past examples of domain fronting, broken trust has not been an issue because the user is a willing participant in the deception. But, if an attacker were to modify the HTTP \texttt{Host} without the victim's knowledge, the end-to-end security guarantees of HTTPS no longer hold.

We have developed a proof-of-concept attack that leverages dynamic linker hijacking \cite{mitre22dllhijacking} through the Linux \texttt{LD\_PRELOAD} trick \cite{goldsborough16ldpreload} to rewrite the HTTP \texttt{Host} field immediately preceding the encryption of the HTTP request. While the proof-of-concept requires an attacker to have a presence on the endpoint, there are other ways to achieve the intended functionality, e.g., by using a supply chain compromise \cite{mitre22supplychain} of popular browsers or TLS libraries.

In the attack, the victim first initiates a TLS handshake with the target domain, and the CDN establishes the TLS connection with the target's proper certificate. When the victim makes an HTTP request, the HTTP \texttt{Host} is overwritten to point to the attacker controlled domain. A CDN that supports domain fronting will then route the HTTP request to the origin server specified by the attacker. The attacker can now view all the decrypted traffic and modify or otherwise censor the decrypted traffic. This behavior clearly violates the authenticity, confidentiality, and integrity guarantees that HTTPS claims to provide. The attacker can optionally create a stealthy command-and-control channel by rewriting request headers and adding HTTP response headers. Unlike simply overriding DNS responses on the endpoint or directly exfiltrating the decrypted data to an attacker-owned server, our attack does not modify the IP address or generate additional network connections, making the attack significantly more stealthy.


\section{Background and Related Work}
\label{sec:background}

The domain name misinformation techniques discussed in Section \ref{sec:ontology} rely on several network protocol standards along with the general mechanics of CDNs, both of which are introduced in this section. We conclude this section by reviewing relevant related work.

\subsection{Network Protocols}

The Domain Name System (DNS) \cite{dnsterminology} provides various mechanisms to associate information with domain names, e.g., it can translate human-readable domain names into routable IP addresses. The extension mechanism around DNS, \texttt{EDNS0} \cite{edns0}, allows for larger message sizes and more advanced handling of DNS requests. In the context of this paper, \texttt{EDNS0} is important because it facilitates DNS responses that optimize the returned IP addresses based on geography in order to reduce latency.

A DNS CNAME record maps an alias domain to the true, canonical name. CNAME records are useful when a single server hosts multiple subdomains, e.g., both \texttt{foo.example.com} and \texttt{bar.example.com} will be aliases of \texttt{example.com}. CNAME records are also useful in the context of CDNs as described below.

After the client obtains an IP address via DNS, the client then begins to directly communicate with the server. For our purposes, we assume the client uses Transport Layer Security (TLS) 1.3 \cite{tls13}. After the TCP handshake, the client sends a TLS \texttt{client\_hello} handshake record specifying its supported cryptographic parameters and supplying some additional data such as the \texttt{server\_name} extension, which provides the domain name of the server the client wishes to communicate with. The \texttt{server\_name} is particularly useful in virtual hosting environments, like CDNs, to help route the connection to the backend server without the CDN's load balancer having to man-in-the-middle the connection.

The server responds with a \texttt{server\_hello} handshake record selecting a set of cryptographic parameters based on the client's preferences. Older versions of TLS negotiate the remainder of the handshake in the clear, while TLS 1.3 begins to encrypt the handshake records.

A TLS 1.3-capable server then sends an encrypted \texttt{certificate} handshake record. This record contains a chain of certificates that allows the client to verify the identity of the server. Wildcard certificates use a wildcard character ($*$) as a subdomain in either the \texttt{subject} or \texttt{subjectAltName} field and allows the certificate to secure multiple subdomains belonging to the same domain name. Finally, the client and server finish the key exchange and begin to exchange \texttt{application\_data} records encrypted with the negotiated keys.

For the purpose of this paper, we assume the client and server exchange encrypted messages using HTTP/1.1 \cite{http1} or HTTP/2 \cite{http2}. The intended server's domain name is located in the \texttt{Host} (HTTP/1.1) or \texttt{:authority} (HTTP/2) field, and the target URI is in the \texttt{Request-Line} (HTTP/1.1) or \texttt{:path} (HTTP/2) field. HTTP/3 \cite{http3} is the latest incarnation of HTTP and it uses the QUIC transport \cite{quic,quictls}. QUIC uses TLS for key negotiation and many of the observations in this paper are directly applicable, but deeper analysis is out-of-scope for the current work.

Because any on-path observer can view and potentially modify plaintext DNS traffic, encrypted DNS protocols were developed to take advantage of the above protocols to secure DNS. DNS-over-HTTPS (DoH) \cite{doh} is one such protocol that maps DNS requests and responses to HTTP and encrypts the connection using TLS. Other encrypted DNS protocols include DNS-over-TLS (DoT) \cite{dot} and DNS-over-QUIC (DoQ) \cite{doq}.

In summary, from the point-of-view of a passive network observer and the above protocols, domain names would appear in the unencrypted DNS request and response, the TLS \texttt{client\_hello} handshake record, and the TLS \texttt{certificate} handshake record for non-1.3 versions of TLS. Domain names are opaque in DoH and TLS 1.3 \texttt{certificate} handshake records.


%
%
%
%
%
%
%
%
%

\subsection{Content Delivery Networks}

Content delivery networks have the goals of reducing latency and improving redundancy for hosted artifacts, while also protecting against security threats, e.g., denial of service attacks. The CDN's servers are geographically dispersed and placed at strategic locations to reduce latency. If a client requests content from a domain that uses a CDN, the DNS response will typically contain a CNAME record where the IP address belongs to the CDN.

The client will then initiate a TLS handshake with the CDN, which hosts the intended domain's certificate. After the TLS handshake, the CDN will proxy the HTTP traffic on behalf of the origin server, which maintains the content requested by the user. If the requested content is not stale, the CDN will return a cached version of the content, and otherwise will request and cache the latest version from the origin server.

\subsection{Related Work}

Fifield et al. \cite{fifield2015blocking} presented the first academic treatment of domain fronting as a censorship circumvention tool. They described 7 CDNs that supported domain fronting at the time of publication. They additionally implemented and studied the deployment of domain fronting in Tor \cite{meek}, Lantern \cite{lantern}, and Psiphon \cite{psiphon}. The authors also examined some detection mechanisms based on network traffic analysis, e.g., packet lengths, and concluded that there were no reliable traffic characteristics that would allow one to detect domain fronting.

Wang et al. \cite{wang2015seeing} studied the ability to detect several network protocol obfuscators including Tor/meek leveraging Google and Amazon for domain fronting. One of their detection strategies included machine learning classifiers that used entropy-based, timing-based, and packet-header data features. Similarly, Li et al. \cite{li2021identification} evaluated a convolutional neural network and features based on packet lengths to detect Tor/meek leveraging Azure and Fastly for domain fronting. Both papers found that a well-resourced censor could reliably detect meek using domain fronting with a low false positive rate. Importantly, Wang et al. note, ``the detection techniques we explore can be, in turn, easily circumvented in almost all cases with simple updates to the obfuscator" \cite{wang2015seeing}.

Instead of viewing domain fronting through the lens of a privacy enhancing technology, Dunwoody \cite{apt2017dunwoody} examined meek/Google domain fronting as it is used by a nation-state attacker, APT29 \cite{mitre22apt29}, in order to evade detection. Similarly, McLellan et al. \cite{smoking21mclellan} investigated how the UNC2465 ransomware group used a legitimate Microsoft domain as a front for their hard-coded domain, \texttt{max-ghoster1.} \texttt{azureedge[.]net}.

\begin{figure}
\centering 
\resizebox{82mm}{!}{
\begin{tikzpicture}

\node (linux_desktop) at (-10,0) {\includegraphics[scale=0.5]{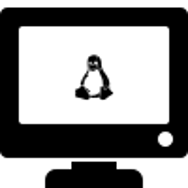}};

\node (dns) at (-4,0) {\includegraphics[scale=0.8]{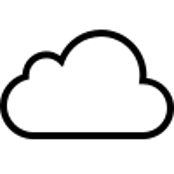}};
\node at (-4,-0.1) {\large DNS};
\node at (-4,1) {\ttfamily\textbf{\textcolor{blue}{a.a.a.a}}};

\node (cdn) at (0,0) {\includegraphics[scale=0.8]{figures/cloud.png}};
\node at (0,-0.1) {\large CDN};
\node at (0,1) {\ttfamily\textbf{\textcolor{blue}{b.b.b.b}}};

\node (origin) at (4,0) {\includegraphics[scale=0.5]{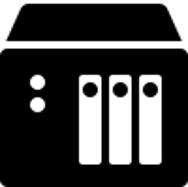}};
\node at (4,1.5) {\ttfamily\textbf{\textcolor{red}{c.c.c.c}}};
\node at (4,1) {\ttfamily\textbf{\textcolor{red}{blocked.org}}};

\draw[line width=.4mm, -] (-10, -1) -- (-10,-8);
\draw[line width=.4mm, -] (-4, -1) -- (-4,-3);
\draw[line width=.4mm, -] (0, -1) -- (0,-8);
\draw[line width=.4mm, -] (4, -1) -- (4,-8);

\node[draw] at (-7.8,-1.65) {\ttfamily\textbf{DNS:} \textbf{\textcolor{blue}{allowed.org}}};
\draw[line width=.4mm, ->] (-10, -1.9) -- (-4,-1.9);

\node[draw] at (-5.8,-2.65) {\ttfamily\textbf{DNS:} \textbf{\textcolor{blue}{b.b.b.b}}};
\draw[line width=.4mm, <-] (-10, -2.9) -- (-4,-2.9);

\draw[line width=.4mm, ->] (-10, -3.9) -- (0,-3.9);
\node[draw] at (-7.35,-3.63) {\ttfamily\textbf{TLS SNI:} \textbf{\textcolor{blue}{allowed.org}}};

\draw[line width=.4mm, <-] (-10, -4.9) -- (0,-4.9);
\node[draw] at (-2.7,-4.63) {\ttfamily\textbf{TLS cert:} \textbf{\textcolor{blue}{allowed.org}}};

\draw[line width=.4mm, ->] (-10, -5.9) -- (0,-5.9);
\node[draw, fill=black!18!white] at (-6.82,-5.63) {\ttfamily\textbf{HTTP Request:} \textbf{\textcolor{red}{blocked.org}}};

\draw[line width=.4mm, ->] (0, -6.4) -- (4,-6.4);
\draw[line width=.4mm, <-] (0, -6.9) -- (4,-6.9);

\draw[line width=.4mm, <-] (-10, -7.4) -- (0,-7.4);
\node[draw, fill=black!18!white] at (-1.9,-7.15) {\ttfamily\textbf{HTTP Response}};

\end{tikzpicture}
}
\caption{A visual representation of domain fronting. Domain name and IP addresses are not meant to be representative of any real-world services. Note that the allowed domain name is visible in DNS and the TLS handshake, and the blocked domain is not visible within the encrypted HTTP request.}
\label{fig:domain-fronting}
\end{figure}

\section{Misinformation Ontology}
\label{sec:ontology}

While we have mainly highlighted domain fronting as a domain name misinformation technique up to this point, there are several related techniques worth noting. In this section, we review the techniques investigated in Section \ref{sec:measurement-results}: domain fronting, domain faking, and domainless fronting. We additionally review two similar techniques that do not cleanly fit the misinformation characterization: wildcard certificate fronting and domain shadowing. For the purposes of this paper, an evasion technique is considered to be domain name misinformation if the encrypted HTTP \texttt{Host} value does not match the TLS \texttt{server\_name} value and is not covered by the TLS certificate's \texttt{subject} or \texttt{SubjectAltName} extension.


\subsection{Domain Fronting}

Domain fronting is the misinformation technique that has garnered the most attention from the privacy research \cite{fifield2015blocking,wang2015seeing} and incident response \cite{apt2017dunwoody,smoking21mclellan} communities. Figure \ref{fig:domain-fronting} illustrates the key features of domain fronting, where the client wishes to communicate with \texttt{blocked.org} by fronting \texttt{allowed.org}. Domain fronting starts with a DNS request for some popular, allowed domain, in this example: \texttt{allowed.org}. This hypothetical domain is hosted by a CDN that supports domain fronting and has an edge device that maps to the \texttt{b.b.b.b} IP address, which the DNS server returns.

The client then initiates a TLS handshake with the CDN and sets the \texttt{client\_hello}'s \texttt{server\_name} to \texttt{allowed.org}. Because the CDN controls the certificates for both of the unrelated domains, \texttt{blocked.org} and \texttt{allowed.org}, it returns the proper certificate for \texttt{allowed.org}. After the TLS handshake, the client sends an encrypted \texttt{application\_data} record with an HTTP request where the HTTP \texttt{Host} field is set to \texttt{blocked.org}. The edge device then decrypts the record and extracts the \texttt{Host} field. Depending on the cache configuration and state, the edge device will then reach out to the origin server of \texttt{blocked.org}, ignoring the value presented in the TLS \texttt{server\_name}, and return the requested content; successfully evading DNS and TLS-layer enforcement of \texttt{blocked.org}.

Domain fronting is possible because the origin server allows the CDN to host their domain's TLS certificate and to decrypt all traffic destined to the origin server. Domain fronting may be an intended feature or an architectural flaw, i.e., the CDN may not keep state associating the \texttt{server\_name} of the original TLS \texttt{client\_hello} with the value present in the HTTP \texttt{Host}. If that state is not present or is purposefully ignored, then the CDN can route the HTTP request at its discretion.

\subsection{Domain Faking}

Domain faking occurs when the server or edge device returns the same certificate and serves content for domains secured by that certificate irrespective of the value present in the TLS \texttt{server\_name}. From the perspective of a passive network observer, domain fronting appears to be legitimate because the edge device returns a valid certificate and serves content for the fronted domain. In contrast, a device supporting domain faking does not return a valid certificate for the fronted domain because it is not authorized to serve content on behalf of the fronted domain. Similarly, a DNS request for the fronted domain will not point to the domain faking device.

For domain faking to appear reasonable to passive network observers, the operator needs to leverage recently developed standards to obfuscate the exchanges in Figure \ref{fig:domain-fronting} that cannot be modified, i.e., the DNS exchange and the TLS certificate. A client begins domain faking by initiating an encrypted DNS request containing the blocked domain, e.g., by using DNS-over-HTTPS \cite{doh}.

The client then sends a TLS 1.3 \texttt{client\_hello} with the \texttt{server\_name} set to the allowed domain. The server is configured to ignore the \texttt{server\_name} extension and returns its default certificate, which is encrypted with TLS 1.3 \cite{tls13}. The client is configured to ignore the returned certificate. The client and server then complete the TLS handshake and begin exchanging encrypted records.

Domain faking is successful for two reasons. First, the same entity has some control over the client and server, which allows it to ignore errors that would otherwise result in a TLS handshake failure. Second, domain faking requires the censor to perform significantly more work in the form of either blocking all encrypted/unsanctioned DNS, scanning the server to retrieve its certificate, or maintaining state that maps IP addresses to domains.

Telegram \cite{telegram} serves as a real-world example of an application that supported domain faking. It used encrypted DNS, primarily to \texttt{dns.google.com}, and a TLS 1.3 handshake. Telegram set the \texttt{server\_name} to \texttt{www.google.com}, but the visited IP addresses are entirely unrelated to Google's infrastructure, e.g., \texttt{5.28.195.163} belongs to the \texttt{CW Vodafone Group PLC} autonomous system. Telegram appears to have deprecated this behavior in November 2022.


\subsection{Domainless Fronting}

Fifield et al. \cite{fifield2015blocking} put forth the concept of domainless fronting, which has many of the same limitations as domain faking. Its main feature is purposefully omitting the \texttt{server\_name} extension or leaving it blank. We consider domainless fronting to misinform when it is used to evade censorship, as opposed to when the \texttt{server\_name} is omitted due to the client using an obsolete version of TLS. Similar to domain faking, it must either use hard-coded IP addresses or rely on encrypted DNS.

TLS 1.3 is helpful to obfuscate the certificate but is not always necessary depending on the hosting infrastructure and configuration. For example, Alibaba's CDN will return a generic certificate with the subject set to \texttt{*.alicdn.com}, which provides little information, but many Akamai-hosted domains will return an informative certificate. Unsurprisingly, this difference is typically related to whether the hosting provider allows domains to be hosted on static IP addresses.

Domainless fronting is appealing because it is relatively simple to configure and does not need to rely on a domain name owned by another entity, reducing the likelihood of collateral damage. Additionally, TLS sessions that naturally omit the \texttt{server\_name} are relatively common but may be trending lower. Fifield et al. reported that 16.5\% of TLS connections lacked the \texttt{server\_name} extension in June 2014 \cite{fifield2015blocking}, Anderson et al. had that number at $\sim$10\% in the first half of 2019 \cite{anderson2019tlsbeyond}, and we observed that number to be 7.7\% in October 2022. While these numbers may not be methodologically comparable, the potential trend is interesting and deserves further investigation.

\subsection{Odds and Ends}


Wildcard certificate fronting takes advantage of the default domain names and certificates provided by CDNs and cloud infrastructure providers to allow fronting between subdomains secured by the default certificate, despite there being no relation between the owners of those subdomains. For example, if you create a public S3 bucket without a custom domain, AWS assigns a generic domain name of the form \texttt{<bucket-name>.s3.amazonaws.com} and provides a default TLS certificate which covers \texttt{*.s3.amazonaws.com} and \texttt{s3.amazonaws.com}. Google Cloud provides similar mechanics for their storage service, \texttt{*.storage.googleapis.com}.

While wildcard certificates are not inherently a security risk in the general case, wildcard certificates that secure many subdomains are worth investigating. This is especially true for the case of subdomains that are unrelated, i.e., the organizations that own and maintain the resources are distinct entities and are only related by the fact that they pay to use the same infrastructure. We further discuss the security risks of wildcard certificate fronting with \texttt{*.cloudfront.net} domains in the context of our novel attack in Section \ref{sec:attacking-df}.

Another related technique is domain shadowing, which was recently introduced by Wei \cite{wei2021domain}. Domain shadowing relies on a popular CDN feature: rewriting the HTTP \texttt{Host} field. The user first registers a new domain that will be used to access blocked resources, \texttt{shadow.com}, then binds that domain to the target domain, \texttt{target.com}, and finally creates a rule to rewrite the \texttt{Host} field from \texttt{shadow.com} to \texttt{target.com} for incoming requests.

Domain shadowing has the clear advantage of the HTTP \texttt{Host} field not indicating the target domain until it is rewritten, which could potentially evade decrypting firewalls. But, for the current topic, we do not consider it misinformation because the shadowed domain has a one-to-one relationship with the target domain and the shadowed domain would match the TLS certificate when the HTTP request arrived at the CDN.

\bgroup
\def\arraystretch{1.02}
\begin{table}
\begin{center}
\begin{tabular}{|l|c|c|c|}
  \hline
  \multicolumn{1}{|c}{Scan} & \multicolumn{1}{|c}{IP Addr} & \multicolumn{1}{|c}{TLS \texttt{server\_name}} & \multicolumn{1}{|c|}{HTTP \texttt{Host}} \\
  \hline
  \hline
  baseline-0 & target\_ip & target\_domain & target\_domain \\
  \hline
  baseline-1 & front\_ip  & front\_domain  & front\_domain  \\
  \hline
  fronting   & front\_ip  & front\_domain  & target\_domain \\
  \hline
  faking     & target\_ip & front\_domain  & target\_domain \\
  \hline
  domainless & target\_ip & -              & target\_domain \\
  \hline
\end{tabular}
\end{center}
\caption{Summary of the 5 executed scans for each pair of destination tuples.}
\label{table:scans}
\end{table}
\egroup

\section{Measuring Domain Name Misinformation}
\label{sec:measurement}

We developed a scanning methodology to detect the domain name misinformation techniques of Section \ref{sec:ontology} and to better understand their support on the Internet. The results are presented in progressively more granular groupings of destinations, first assessing an autonomous system-wide grouping of destinations and finishing with a grouping based on the fully qualified domain name (FQDN) of the DNS canonical name. As we will show, much of the ambiguity around organizations' support for these techniques erodes with more specific groupings. The last subsection finishes by discussing the prevalence of this support among the most popular domains.

\bgroup
\def\arraystretch{1.02}
\begin{table*}
\begin{center}
\begin{tabular}{|l|r|r|r|r|}
  \hline
  \multicolumn{1}{|c|}{\multirow{2}{*}{Autonomous System}} & \multicolumn{1}{c|}{Number of} & \multicolumn{1}{c|}{Domain Fronting} & \multicolumn{1}{c|}{Domain Faking} & \multicolumn{1}{c|}{Domainless Fronting} \\
  \multicolumn{1}{|c|}{} & \multicolumn{1}{c|}{Observed Domains} & \multicolumn{1}{c|}{Support} & \multicolumn{1}{c|}{Support} & \multicolumn{1}{c|}{Support} \\
  \hline
  \hline
  ACE                                  &    387 &  \cellcolor{red!25}91.45\% & \cellcolor{red!25}100.00\% &  \cellcolor{red!25}99.15\% \\
  \hline
  AKAMAI-AS                            &  4,612 &   \cellcolor{green!25}4.44\% &  \cellcolor{red!25}99.38\% &  \cellcolor{red!25}99.32\% \\
  \hline
  Akamai International B.V.            &  4,169 &   \cellcolor{green!25}2.55\% &  \cellcolor{yellow!25}56.06\% &  \cellcolor{yellow!25}56.11\% \\
  \hline
  AMAZON-02                            & 16,841 &   \cellcolor{green!25}4.97\% &  \cellcolor{yellow!25}45.30\% &  \cellcolor{yellow!25}55.32\% \\
  \hline
  AMAZON-AES                           &  9,437 &   \cellcolor{green!25}4.02\% &  \cellcolor{yellow!25}69.27\% &  \cellcolor{yellow!25}75.33\% \\
  \hline
  Beijing Baidu Netcom                 &    186 &  \cellcolor{yellow!25}17.39\% &  \cellcolor{red!25}97.67\% &  \cellcolor{red!25}97.67\% \\
  \hline
  Chinanet                             &  1,565 &  \cellcolor{yellow!25}34.48\% &  \cellcolor{yellow!25}86.02\% &  \cellcolor{red!25}95.75\% \\
  \hline
  CLOUDFLARENET                        & 24,025 &   \cellcolor{green!25}0.00\% &   \cellcolor{green!25}0.77\% &   \cellcolor{yellow!25}6.88\% \\
  \hline
  Datacamp Limited                     &  1,259 &  \cellcolor{yellow!25}73.00\% &  \cellcolor{red!25}99.61\% & \cellcolor{red!25}100.00\% \\
  \hline
  DIGITALOCEAN-ASN                     &  3,837 &   \cellcolor{green!25}2.07\% &  \cellcolor{yellow!25}84.70\% &  \cellcolor{red!25}96.79\% \\
  \hline
  EDGECAST                             &  1,260 &  \cellcolor{yellow!25}83.59\% & \cellcolor{red!25}100.00\% & \cellcolor{red!25}100.00\% \\
  \hline
  FASTLY                               &  5,874 &  \cellcolor{yellow!25}62.62\% &  \cellcolor{red!25}98.76\% &  \cellcolor{red!25}98.61\% \\
  \hline
  GOOGLE-CLOUD-PLATFORM                &  6,471 &  \cellcolor{yellow!25}12.82\% &  \cellcolor{yellow!25}88.53\% &  \cellcolor{yellow!25}90.94\% \\
  \hline
  Hangzhou Alibaba Advertising Co.,Ltd &  2,772 &   \cellcolor{yellow!25}7.66\% &  \cellcolor{yellow!25}93.57\% &  \cellcolor{red!25}97.86\% \\
  \hline
  INCAPSULA                            &  1,872 & \cellcolor{red!25}100.00\% & \cellcolor{red!25}100.00\% &  \cellcolor{red!25}98.05\% \\
  \hline
  LLNW                                 &    157 &  \cellcolor{yellow!25}91.53\% & \cellcolor{red!25}100.00\% & \cellcolor{red!25}100.00\% \\
  \hline
  MICROSOFT-CORP-MSN-AS-BLOCK          &  6,867 &   \cellcolor{green!25}3.31\% &  \cellcolor{yellow!25}88.29\% &  \cellcolor{yellow!25}94.72\% \\
  \hline
  OVH SAS                              &  4,426 &   \cellcolor{green!25}2.88\% &  \cellcolor{yellow!25}74.26\% &  \cellcolor{red!25}97.55\% \\
  \hline
  QUANTILNETWORK                       &    298 &  \cellcolor{yellow!25}82.62\% &  \cellcolor{red!25}98.32\% &  \cellcolor{red!25}98.66\% \\
  \hline
  STACKPATH-CDN                        &  1,380 & \cellcolor{red!25}100.00\% & \cellcolor{red!25}100.00\% & \cellcolor{red!25}100.00\% \\
  \hline
  Wix.com Ltd.                         & 12,248 & \cellcolor{red!25}100.00\% & \cellcolor{red!25}100.00\% &   \cellcolor{green!25}1.21\% \\
  \hline
  Zhejiang Taobao Network Co.,Ltd      &  1,360 & \cellcolor{red!25}100.00\% & \cellcolor{red!25}100.00\% & \cellcolor{red!25}100.00\% \\
  \hline
\end{tabular}
\end{center}
\caption{Scan results when using autonomous systems to group destinations. The color legend is as follows: Green ($<$ 5\%), Yellow (5\% $\geq$ and $<$ 95\%), and Red ($\geq$ 95\%)}
\label{table:scan-results-as}
\end{table*}
\egroup

\subsection{Methodology}
\label{sec:measurement-methodology}

Given a pair of related (\texttt{domain}, \texttt{ip})-tuples, one marked as the target and the other as the front, our scanning system is designed to identify if the techniques described in Section \ref{sec:ontology} apply. The relationships examined in Section \ref{sec:measurement-results} are based on autonomous systems, and the domain name and FQDN of the canonical name present in the DNS CNAME record. The system generates a candidate set of related destination tuples by grouping them based on these criteria.

For each destination tuple, the scanner generates 5 pairs of tuples by randomly selecting other destinations in the candidate set. Table \ref{table:scans} describes the 5 scans that are executed given a pair of destination tuples, where the target is (\texttt{target\_ip}, \texttt{target\_domain}) and the front is (\texttt{front\_ip}, \texttt{front\_domain}). For each scan, Table \ref{table:scans} lists the IP address, TLS \texttt{server\_name}, and encrypted HTTP \texttt{Host} used for the connection.

\texttt{baseline-0} and \texttt{baseline-1} are used to determine if the misinformation techniques were successful and simply scan the target and front IP/domain, respectively. The three remaining scans attempt to use the two destinations to perform domain fronting, domain faking, and domainless fronting, setting the protocol fields as given in Table \ref{table:scans}.

To determine if the misinformation techniques were successful, the scanner collects several response features for each scan:
\begin{itemize}
\item a JSON representation of the TLS certificate
\item the HTTP \texttt{status\_code}
\item the full list of HTTP response headers and values
\item the length of the returned content
\end{itemize}
If either of the TLS certificates associated with the baseline scans secures both domains, e.g., both domains appear in the \texttt{subjectAltName} extension, domain fronting is not applicable, and the results are ignored. If either of the baseline scans returns a non-\texttt{200} HTTP \texttt{status\_code}, the results are also ignored. This pruning may be overly aggressive, but it helps to remove ambiguity, which made analyzing the results more straightforward.

After the above pruning, the analysis considers the non-baseline scans to be successful if the scan:
\begin{enumerate}
\item returns a \texttt{200} HTTP \texttt{status\_code}, and
\item the length of the returned content matches that of \texttt{baseline-0} but not \texttt{baseline-1}.
\end{enumerate}
To better handle dynamic content, the system makes two exceptions to the second criteria. First, if the length of the returned content for a misinformation scan is within 5\% of \texttt{baseline-0}'s length and not within 20\% of \texttt{baseline-1}'s length, it is considered successful. All manually inspected instances of dynamic content satisfying this exception were correct. Second, if the HTTP response header names and ordering for a misinformation scan exactly matches those of \texttt{baseline-0} and not \texttt{baseline-1}, it is considered successful. Roughly 87\% of the successful misinformation scans maintained ordering for HTTP response headers. In most failure cases, the server responds with either \texttt{403 Forbidden}, \texttt{421 Misdirected Request}, or \texttt{400 Bad Request}.



The scanning code was written in Python and PySpark and was deployed on an Amazon EMR cluster with 300 executors. The scanner uses the Python \texttt{requests} library to make connections and specified the following HTTP headers for each scan:
\lstset{
  basicstyle=\small\ttfamily,
  breaklines=true,
  escapeinside=||
}
\begin{lstlisting}
  headers  = {'Host':       http_host,
              'User-Agent': USER_AGENT,
              'Connection': 'close'}
\end{lstlisting}
where \texttt{http\_host} is given in Table \ref{table:scans} and \texttt{USER\_AGENT} described a Chrome 104 client running on Windows 10.

To have more control over the destination IP address, the scanner uses a monkey patch for \texttt{socket.getaddrinfo} before each scan that hardcodes the returned IP address:
\begin{lstlisting}
socket.getaddrinfo = (lambda *args:
        [(socket.AddressFamily.AF_INET,
          socket.SocketKind.SOCK_STREAM,
          6, '', (dst_ip, 443))])
\end{lstlisting}
where \texttt{dst\_ip} is given in Table \ref{table:scans}.

To make the scanning more efficient, the system runs a prefiltering scan immediately before grouping destination tuples and running the scans in Table \ref{table:scans}. The system scans each destination and filters destinations that do not return a \texttt{200} HTTP \texttt{status\_code}. Despite this filtering step, some baseline scans would return a non-\texttt{200} code. Most of these cases were explained by distributed denial-of-service protections, unsurprising given the parallel nature of the PySpark scanning infrastructure. The system re-ran these failed scans sequentially with a pure Python scanner, and applied the rules listed above to the results.

Details of the specific datasets for each experiment are given in the subsections of Section \ref{sec:measurement-results}. To find the initial set of labels, we analyzed passive DNS data collected from $\sim$80 geographically dispersed sites all belonging to a single multinational enterprise.

The passive DNS data was filtered to only include DNS CNAME records. We then grouped alias domain names by the canonical name's domain name and sorted the canonical name by the number of unique alias domain names that map to it. The most common domains found with this method are reported in Section \ref{sec:measurement-results}. We omitted some domains that were related and behaved similarly, e.g., Edgecast has a series of canonical names that begin with a Greek letter and end in \texttt{cdn.net}, but we only report results for \texttt{*.omicroncdn.net}.

Given the list of domain names associated with canonical names, we generated both more generic and more specific labels. For the more generic autonomous system labels, we collected all IP addresses in the DNS CNAME records for a given canonical name and mapped those IP addresses to their autonomous systems. We report results for the two most prevalent autonomous systems for each canonical name, which covered almost all observed records.

For the more specific canonical name FQDNs, we performed an analysis similar to that of the canonical name's domain name, i.e., we grouped all alias domains by canonical name FQDNs and sorted the FQDNs by their number of unique alias domain names. Some CDNs like Baidu and Fastly have a relatively well-defined, small set of canonical names that service a large number of distinct customers. Other CDNs, like Cloudflare and StackPath, often encode a customer-specific domain or a unique ID as a subdomain of the canonical name. The FQDN measurement also explains the ambiguity of misinformation support when looking at more generic relationships.

\bgroup
\def\arraystretch{1.02}
\begin{table*}
\begin{center}
\begin{tabular}{|l|r|r|r|r|r|r|}
  \hline
  \multicolumn{1}{|c|}{\multirow{2}{*}{DNS CNAME Domain}} & \multicolumn{1}{|c|}{Number of} & \multicolumn{1}{c|}{Domain Fronting} & \multicolumn{1}{c|}{Domain Faking} & \multicolumn{1}{c|}{Domainless Fronting} \\
  \multicolumn{1}{|c|}{} & \multicolumn{1}{|c|}{Observed Domains} & \multicolumn{1}{c|}{Support} & \multicolumn{1}{c|}{Support} & \multicolumn{1}{c|}{Support} \\
  \hline
  \hline
  *.akamaiedge.net. (Akamai)         & 58,821 &   \cellcolor{green!25}1.65\% &  \cellcolor{yellow!25}40.47\% &  \cellcolor{yellow!25}39.64\% \\
  \hline
  *.cdngslb.com. (Alibaba)           &  4,961 &  \cellcolor{red!25}99.80\% &  \cellcolor{red!25}98.96\% &  \cellcolor{red!25}99.69\% \\
  \hline
  *.kunlunar.com. (Alibaba)          &  1,330 & \cellcolor{red!25}100.00\% & \cellcolor{red!25}100.00\% & \cellcolor{red!25}100.00\% \\
  \hline
  *.cloudfront.net. (AWS)         & 92,369 &   \cellcolor{green!25}2.72\% &   \cellcolor{green!25}1.89\% &   \cellcolor{green!25}0.99\% \\
  \hline
  *.amazonaws.com. (AWS)          & 43,777 &  \cellcolor{yellow!25}14.35\% &  \cellcolor{yellow!25}94.96\% &  \cellcolor{yellow!25}74.72\% \\
  \hline
  *.jomodns.com. (Baidu)             &  1,462 &  \cellcolor{yellow!25}94.51\% & \cellcolor{red!25}100.00\% & \cellcolor{red!25}100.00\% \\
  \hline
  *.b-cdn.net. (Bunny CDN)           &  4,135 & \cellcolor{red!25}100.00\% & \cellcolor{red!25}100.00\% & \cellcolor{red!25}100.00\% \\
  \hline
  *.cdn77.org. (CDN77)               &  1,565 & \cellcolor{red!25}100.00\% & \cellcolor{red!25}100.00\% & \cellcolor{red!25}100.00\% \\
  \hline
  *.cloudflare.net. (Cloudflare)     & 67,170 &   \cellcolor{green!25}0.00\% &   \cellcolor{green!25}0.53\% &  \cellcolor{yellow!25}11.69\% \\
  \hline
  *.ovscdns.com. (DNStination)       &  2,697 &  \cellcolor{yellow!25}92.43\% & \cellcolor{red!25}100.00\% &  \cellcolor{red!25}99.71\% \\
  \hline
  *.omicroncdn.net. (Edgecast)       &    653 &  \cellcolor{red!25}98.02\% & \cellcolor{red!25}100.00\% & \cellcolor{red!25}100.00\% \\
  \hline
  *.edgecastcdn.net. (Edgecast)      &  1,125 &  \cellcolor{yellow!25}87.79\% & \cellcolor{red!25}100.00\% & \cellcolor{red!25}100.00\% \\
  \hline
  *.fastly.net. (Fastly)             & 56,609 &  \cellcolor{yellow!25}72.09\% &  \cellcolor{red!25}99.31\% &  \cellcolor{red!25}99.30\% \\
  \hline
  *.googlehosted.com. (Google)       &  7,640 & \cellcolor{red!25}100.00\% & \cellcolor{red!25}100.00\% &   \cellcolor{green!25}0.00\% \\
  \hline
  *.google.com. (Google)             & 12,894 &  \cellcolor{yellow!25}65.45\% &  \cellcolor{red!25}99.97\% &   \cellcolor{yellow!25}5.36\% \\
  \hline
  *.impervadns.net. (Imperva)        &  5,118 &  \cellcolor{red!25}99.82\% &  \cellcolor{red!25}99.10\% &  \cellcolor{red!25}96.60\% \\
  \hline
  *.incapdns.net. (Imperva)          &  9,646 &  \cellcolor{red!25}99.73\% &  \cellcolor{red!25}98.99\% &  \cellcolor{red!25}97.28\% \\
  \hline
  *.kxcdn.com. (KeyCDN)              &  1,526 & \cellcolor{red!25}100.00\% & \cellcolor{red!25}100.00\% & \cellcolor{red!25}100.00\% \\
  \hline
  *.llnwi.net. (Limelight)           &    377 & \cellcolor{red!25}100.00\% & \cellcolor{red!25}100.00\% & \cellcolor{red!25}100.00\% \\
  \hline
  *.fdv2-t-msedge.net. (Microsoft)   & 12,077 &  \cellcolor{yellow!25}91.96\% &  \cellcolor{yellow!25}91.91\% & \cellcolor{red!25}100.00\% \\
  \hline
  *.cloudapp.net. (Microsoft)        & 24,401 &   \cellcolor{green!25}1.75\% &  \cellcolor{red!25}95.00\% &  \cellcolor{red!25}95.05\% \\
  \hline
  *.netlifyglobalcdn.com. (Netlify)  &    499 & \cellcolor{red!25}100.00\% & \cellcolor{red!25}100.00\% & \cellcolor{red!25}100.00\% \\
  \hline
  *.netlify.com. (Netlify)           &    923 & \cellcolor{red!25}100.00\% & \cellcolor{red!25}100.00\% & \cellcolor{red!25}100.00\% \\
  \hline
  *.stackpathcdn.com. (StackPath)    &  4,559 & \cellcolor{red!25}100.00\% & \cellcolor{red!25}100.00\% & \cellcolor{red!25}100.00\% \\
  \hline
  *.hwcdn.net. (StackPath)           &  2,466 & \cellcolor{red!25}100.00\% & \cellcolor{red!25}100.00\% & \cellcolor{red!25}100.00\% \\
  \hline
  *.vercel-dns.com. (Vercel)         &  4,767 & \cellcolor{red!25}100.00\% & \cellcolor{red!25}100.00\% & \cellcolor{red!25}100.00\% \\
  \hline
  *.wswebpic.com. (Wangsu)           &  1,385 &  \cellcolor{yellow!25}78.96\% &  \cellcolor{red!25}98.44\% &  \cellcolor{red!25}97.66\% \\
  \hline
  *.wswebcdn.com. (Wangsu)           &    897 &  \cellcolor{yellow!25}85.25\% &  \cellcolor{red!25}96.30\% &  \cellcolor{red!25}96.63\% \\
  \hline
  *.wixdns.net. (Wix)                & 73,402 & \cellcolor{red!25}100.00\% & \cellcolor{red!25}100.00\% &   \cellcolor{green!25}0.00\% \\
  \hline
\end{tabular}
\end{center}
\caption{Scan results when using the domains in canonical names to group destinations.}
\label{table:scan-results-dn}
\end{table*}
\egroup

\subsection{Results}
\label{sec:measurement-results}

In this section, we present the results of scanning Internet infrastructure to determine support for domain name misinformation. All destination tuples used for scanning in these subsections were collected from the same $\sim$80 geographically dispersed sites belonging to the single multinational enterprise mentioned above, but the type of data varies as discussed below. We used our open-source tool, \texttt{mercury} \cite{mercury}, to collect the necessary network metadata, which was collected between January 22nd, 2023 and February 21st, 2023, and is referred to as \texttt{enterprise-0}.

\subsubsection{Autonomous Systems}

To generate the list of destinations to scan, we extract the TLS \texttt{server\_name} value and destination IP address from all passively observed packets containing a TLS \texttt{client\_hello} from \texttt{enterprise-0}. We then map the IP addresses to their respective autonomous systems, group the destination tuples based on their autonomous system, and filter data that does not belong to a tracked autonomous system. For efficiency reasons, we keep the 100,000 most prevalent destination tuples per autonomous system.

These scan results are presented in Table \ref{table:scan-results-as}. The number of observed domains is after the filtering step described in Section \ref{sec:measurement-methodology}. The cells are highlighted to indicate how much ambiguity there is in the results. Green/red is used when there is less than 5\% or greater than 95\% support for a given technique. Yellow is used for the remaining range and indicates competing architectures within the same autonomous system, where only some support the misinformation technique. For example, only 62.62\% of the domain fronting scans for destinations that map to the \texttt{FASTLY} autonomous system were successful, despite Fastly being known to support domain fronting. We investigate this discrepancy and similar issues in the following subsections.


The relatively high support for domain faking and domainless fronting among most autonomous systems is partly explained by server autonomy, i.e., many of the domains are not associated with a CDN and are responsible for their own server configurations. In this case, most of the IP addresses only map to a single domain.

Unfortunately, there is not always an obvious explanation for the support numbers in Table \ref{table:scan-results-as} because grouping destinations by autonomous system conceals a large amount of diversity in the underlying infrastructure, which motivates the following sections.

\subsubsection{DNS CNAME Domain}

To address some of the limitations from the previous subsection, we now investigate domain name misinformation support when we group destination tuples based on their canonical name's domain name. We first collect all DNS CNAME records from the \texttt{enterprise-0} dataset, and then extract the alias domain, canonical domain, and IP address of the canonical name from each record. Again, we keep the 100,000 most prevalent destination tuples per canonical domain.

The results in Table \ref{table:scan-results-dn} seem to be converging towards more clear answers, but there still exists caveats. Some of the support numbers slightly less than 100\% for a given technique are explained by network failures in the scanning, but the support close to zero is more difficult to explain without intimate knowledge of the platforms.

\texttt{*.cloudflare.net.}'s modest support for domainless fronting is also surprising given Cloudflare's dependence on the \texttt{server\_name} extension. After investigating specific examples, the primary pattern was related to destinations that were hosted on autonomous systems not related to Cloudflare but did use a \texttt{*.cloudflare.net.} canonical name in their DNS records. Microsoft's \texttt{*.cloudapp.net.} will sometimes support domain fronting and faking when the domains map to the same IP address. In these cases, they appear to share the same load balancer, but do have unrelated TLS certificates.

\bgroup
\def\arraystretch{1.02}
\begin{table*}
\begin{center}
\begin{tabular}{|l|r|r|r|r|r|r|}
  \hline
  \multicolumn{1}{|c|}{\multirow{2}{*}{DNS CNAME FQDN}} & \multicolumn{1}{|c|}{Number of} & \multicolumn{1}{c|}{Domain Fronting} & \multicolumn{1}{c|}{Domain Faking} & \multicolumn{1}{c|}{Domainless Fronting} \\
  \multicolumn{1}{|c|}{} & \multicolumn{1}{|c|}{Observed Domains} & \multicolumn{1}{c|}{Support} & \multicolumn{1}{c|}{Support} & \multicolumn{1}{c|}{Support} \\
  \hline
  \hline
  [customer].dscx.akamaiedge.net. (Akamai)                 &  1,099 &   \cellcolor{green!25}0.00\% &   \cellcolor{green!25}0.00\% &   \cellcolor{green!25}0.00\% \\
  \hline
  tinyglobalcdnweb.[xxx].cdngslb.com. (Alibaba)            &    119 & \cellcolor{red!25}100.00\% & \cellcolor{red!25}100.00\% & \cellcolor{red!25}100.00\% \\
  \hline
  globalcdnweb.[xxx].kunlunar.com. (Alibaba)               &    143 & \cellcolor{red!25}100.00\% & \cellcolor{red!25}100.00\% & \cellcolor{red!25}100.00\% \\
  \hline
  [multiscreensites].elb.us-east-1.amazonaws.com. (AWS) &  8,993 & \cellcolor{red!25}100.00\% & \cellcolor{red!25}100.00\% &   \cellcolor{green!25}0.00\% \\
  \hline
  [wshopon].elb.us-east-2.amazonaws.com. (AWS)          &    248 & \cellcolor{red!25}100.00\% & \cellcolor{red!25}100.00\% & \cellcolor{red!25}100.00\% \\
  \hline
  opencdn.jomodns.com. (Baidu)                             &    377 & \cellcolor{red!25}100.00\% & \cellcolor{red!25}100.00\% & \cellcolor{red!25}100.00\% \\
  \hline
  sni1gl.wpc.omicroncdn.net. (Edgecast)                    &    119 & \cellcolor{red!25}100.00\% & \cellcolor{red!25}100.00\% & \cellcolor{red!25}100.00\% \\
  \hline
  j.sni.global.fastly.net. (Fastly)                        &  1,877 & \cellcolor{red!25}100.00\% & \cellcolor{red!25}100.00\% & \cellcolor{red!25}100.00\% \\
  \hline
  ghs.google.com (Google)                                  &  2,747 & \cellcolor{red!25}100.00\% & \cellcolor{red!25}100.00\% &   \cellcolor{green!25}0.00\% \\
  \hline
  www3.l.google.com. (Google)                              &    280 & \cellcolor{red!25}100.00\% & \cellcolor{red!25}100.00\% & \cellcolor{red!25}100.00\% \\
  \hline
  part-xxx.t-xxx.fdv2-t-msedge.net. (Microsoft)            &  1,311 &  \cellcolor{yellow!25}91.21\% &  \cellcolor{yellow!25}91.15\% & \cellcolor{red!25}100.00\% \\
  \hline
  waws-prod-xxx-xxx.cloudapp.net. (Microsoft)              &    165 & \cellcolor{red!25}100.00\% & \cellcolor{red!25}100.00\% & \cellcolor{red!25}100.00\% \\
  \hline
  cname.vercel-dns.com. (Vercel)                           &  4,687 & \cellcolor{red!25}100.00\% & \cellcolor{red!25}100.00\% & \cellcolor{red!25}100.00\% \\
  \hline
  td-ccm- [...] .wixdns.net. (Wix)                         & 36,336 & \cellcolor{red!25}100.00\% & \cellcolor{red!25}100.00\% &   \cellcolor{green!25}0.00\% \\
  \hline
\end{tabular}
\end{center}
\caption{Scan results when using the fully qualified canonical names to group destinations.}
\label{table:scan-results-fqdn}
\end{table*}
\egroup

Viewing the scanning results based on canonical domains does make the hosting providers' support for different misinformation techniques very clear in some cases, e.g., alias domains mapping to \texttt{*.googlehosted.com.} and \texttt{*.wixdns.net.} will support domain fronting and faking, but will not support domainless fronting. But some popular canonical domains remain unclear, e.g., \texttt{*.akamaiedge.net.}, \texttt{*.amazonaws.com.}, and \texttt{*.} \texttt{google.com.}.



\subsubsection{DNS CNAME FQDN}

We are again addressing the previous subsection's limitations by using a more granular grouping of destinations, the canonical name's FQDN. The data preparation was the same as the previous subsection's except for using the canonical name's FQDN.

Table \ref{table:scan-results-fqdn} presents the results, where some of the FQDNs have been shortened if it appeared that they were specific to a customer as opposed to specific to the provider. Support, or lack thereof, for the different misinformation techniques has become much clearer. For instance, \texttt{*.google.com.}'s uncertainty for domain fronting support can now be explained by prohibiting fronting between customer-owned properties (\texttt{ghs.google.com.}) and Google-owned properties (\texttt{www3.l.google.com.}) but allowing fronting within those groups.

Like the above observation, Fastly's mixed support for domain fronting when using \texttt{*.fastly.net.} to group destinations is explained by looking at the more specific canonical FQDNs. For example, we found that domain fronting, domain faking, and domainless fronting are all possible between alias domains that map to the \texttt{j.sni.global.fastly.net.} canonical FQDN. On the other hand, there were some \texttt{*.fastly.net.} canonical FQDNs that contained customer-specific identifiers as subdomains that did not support any of the misinformation techniques.

\texttt{*.amazonaws.com.} is interesting due to the large number of distinct services that are mapped to it, including third-party groups that take advantage of Amazon's Elastic Load Balancing (ELB) service. The two \texttt{*.amazonaws.com.} FQDNs in Table \ref{table:scan-results-fqdn} are companies that use ELB to host their customers' websites. Their backend environments must be configured differently because of the discrepancy in domainless fronting.

Microsoft's \texttt{fdv2-t-msedge.net.} was the one outlier whose domain name misinformation support isn't clear. In November 2022, Azure began to prohibit domain fronting for newly created Azure Front Door resources and created a support system for customers to prohibit domain fronting on older resources \cite{azure22domainfronting}. Azure will discontinue domain fronting for all resources in November 2023. Table \ref{table:scan-results-fqdn} provides a snapshot of this policy's effects, and Azure's fronting support should converge to 0 in November 2023.







\subsection{Discussion}

Using candidate sets based on the CNAME FQDN clearly provides the most information, but these sets are limited because they are not always applicable. As discussed above, many popular canonical FQDNs are specific to a single customer, in which case candidate sets constructed through the canonical domain name would be more informative. Candidate sets based on autonomous systems or subnets are useful when CNAME records do not exist for a given domain.

While we do know the possibilities and constraints with respect to misinformation techniques for some popular canonical names, there are many caveats that deserve further attention. For example, a small set of Akamai-hosted domains will allow domain fronting, which may be related to custom software stacks like Drupal \cite{drupal}, which we observed running US government sites that allow domain fronting. In any case, the complexity of this type of analysis is likely to continue to grow with the complexity of CDNs and deserves further research.

Domain fronting is in part successful due to its selection of a popular domain to use as a front. To better understand popular domains' support for domain fronting, we analyze the freely available Umbrella Popularity List \cite{umbrellatop1m}, which lists the top-1 million queried domains based on their global DNS infrastructure. Using the popularity list from February 21st, 2023, we ran \texttt{dig} on each domain from AWS EC2 instances in the \texttt{us-east-1} and \texttt{ap-southeast-1} regions. We used the default DNS resolver and \texttt{mercury} \cite{mercury} to collect all DNS responses.

\begin{figure*}
\centering 
\resizebox{115mm}{!}{
\begin{tikzpicture}

\node (firefox) at (-10,0) {\includegraphics[scale=0.7]{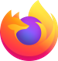}};

\node (cloudfront) at (0,0) {\includegraphics[scale=0.6]{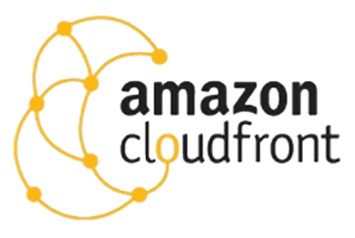}};
\node at (0,-0.8) {\ttfamily\scriptsize \textbf{*.cloudfront.net}};

\node (proxy) at (4,0) {\includegraphics[scale=0.4]{figures/server.png}};
\node at (4,-0.8) {\ttfamily\scriptsize \textbf{\textcolor{red}{ec2-d1.amazonaws.com}}};
\node at (4,-1.12) {\scriptsize \textbf{(reverse proxy)}};

\node (origin) at (4,-3) {\includegraphics[scale=0.4]{figures/server.png}};
\node at (4,-3.8) {\ttfamily\scriptsize \textbf{ec2-d0.amazonaws.com}};

\node[draw] at (-10.1,1.81) {\large Endpoint};

\node[draw, fill=white] at (-8,0) {\large intercept};
\node[draw, fill=white] at (-6,0) {\large libnspr4};

\node at (-8.5,1.1) {\ttfamily\scriptsize \textbf{Host: d0.cloudfront.net}};

\node at (-4.3,1.1) {\ttfamily\scriptsize \textbf{TLS: d0.cloudfront.net}};
\node[draw, fill=black!18!white] at (-4.4,0.7) {\ttfamily\scriptsize \textbf{HTTP: \textcolor{red}{d1.cloudfront.net}}};
\node at (-4.85,-0.65) {\ttfamily\scriptsize \textbf{Content: ec2-d0.amazonaws.com}};

\node at (-1.5,2.8) {\small Also include original Host};
\node[draw, fill=black!18!white] at (0.5,2.3) {\ttfamily\scriptsize \textbf{User-Agent: d0.cloudfront.net...}};

  \begin{pgfonlayer}{background}

	\draw [draw=black] (-11, 1.5) rectangle (-2.5,-1.5);

	\draw[line width=.4mm, ->] (-9.3, 0.1) -- (-1,0.1);
	\draw[line width=.4mm, <-] (-9.3, -0.1) -- (-1,-0.1);

	\draw[line width=.4mm, ->] (1, 0.1) -- (3.3,0.1);
	\draw[line width=.4mm, <-] (1, -0.1) -- (3.3,-0.1);

	\draw[line width=.4mm, ->] (3.9, -1.35) -- (3.9,-2.4);
	\draw[line width=.4mm, <-] (4.1, -1.35) -- (4.1,-2.4);

	\draw[line width=.4mm, -] (-2.6, 0.7) -- (-2.3, 0.7);
	\draw[line width=.4mm, -] (-2.3, 0.7) -- (-2.3, 2.3);
	\draw[line width=.4mm, ->] (-2.3, 2.3) -- (-2, 2.3);

  \end{pgfonlayer}

\end{tikzpicture}
}
\caption{Proof-of-concept attack against \texttt{[*].cloudfront.net} domains. In response to a preprint of this paper, AWS quickly implemented a fix to prevent this attack, but the same concepts apply to other cloud vendors supporting domain fronting.}
\label{fig:attack-poc}
\end{figure*}

331 thousand domains from the top-1 million list mapped to a canonical name. 133 thousand of the domains mapped to a canonical name that was analyzed in Table \ref{table:scan-results-dn}. Using the domain fronting support from Table \ref{table:scan-results-dn}, we estimate that at least 19.4\% of these 133 thousand and 2.5\% of the top-1 million domains can be used as a front.

The DNS resolutions across regions were relatively stable, perhaps unsurprising given the global presence of most CDNs. But there were minor differences between the two regions that may be explained by the need for redundancy and load balancing configurations as opposed to geography. The most notable example being \texttt{ctldl.windowsupdate.com}, which was ranked 14th in the Umbrella list and is used by Microsoft to update its list of trusted and untrusted root certificates. The canonical name returned in the \texttt{us-east-1} region was related to Azure, but it was related to Limelight in the \texttt{ap-southeast-1} region. Within our datasets, we observed the \texttt{ctldl.windowsupdate.com} alias domain mapped to canonical names belonging to Azure, Akamai, Limelight, and StackPath, where the latter two CDNs have broad support for domain fronting.


\section{Exploiting Domain Name Misinformation}
\label{sec:attacking-df}

As we have shown in the previous section, domain fronting support among many of the popular CDNs is not particularly rare. We now shift the focus of this paper towards how a well-resourced adversary can exploit domain fronting to break the end-to-end security guarantees of HTTPS and stealthily man-in-the-middle network sessions that connect to a CDN supporting domain fronting. We implemented a proof-of-concept attack against Fastly and AWS \texttt{[*].cloudfront.net} domains, which will succeed against any of the providers that we report as supporting domain fronting.

Our attack injects a small module into the victim's browser to rewrite selected \texttt{Host} fields, and then greatly amplifies its effect by implementing the proxying and monitoring functions externally. Our proof-of-concept uses dynamic linker hijacking \cite{mitre22dllhijacking}, but many other techniques could be used, including the more generic execution flow hijacking \cite{mitre22hijackexecutionflow}, process injection \cite{mitre22processinjection}, or a supply chain compromise \cite{mitre22supplychain}.

We note that in response to a preprint of this paper, AWS promptly implemented a fix to prevent this attack and domain fronting between CloudFront distribution points is no longer possible.

\subsection{Threat Model}

The main goals of the attacker are two-fold: 1) perform a man-in-the-middle attack to catalog or censor a victim's targeted traffic, where ``targeted" is defined as traffic destined to interesting sites hosted by CDNs that support domain fronting, e.g., \texttt{www.reddit.com}, and 2) establish a stealthy command-and-control channel through HTTP header manipulation.


To execute the attack, the attacker only needs to have the ability to 1) modify HTTP headers either through hijacking the execution flow of a process or by successfully executing a supply chain attack, and 2) create an account on a CDN that supports domain fronting. While the first point is far more onerous than the second, we note that dynamic linker hijacking \cite{mitre22dllhijacking} is a standard attack technique not uncommon in malicious software and is facilitated by attack tools like metasploit \cite{metasploit}.

The attacker does not have the ability to direct the victim towards domains that can be intercepted. If the attacker's goal was to man-in-the-middle \texttt{www.reddit.com} traffic, the victim's traffic destined to \texttt{www.reddit.com} would need to arise organically through the victim's actions. From the point-of-view of establishing a command-and-control channel, this may not be a restriction in practice because a small set of popular domains are visited hundreds of times per day by a given user. In the \texttt{enterprise-0} dataset, \texttt{*.cloudfront.net} domains were visited $\sim$27 times per day per user.

The primary benefits of this attack to the adversary include:
\begin{enumerate}
\itemsep0em
\item There is no need for a separate channel to perform data exfiltration or command-and-control. As is common in remediation efforts, incident responders heavily rely on IP address and domain name-based indicators of compromise to identify infected hosts (e.g., see \texttt{log4shell} \cite{taloslog4shell}), which are not present.
\item From the point-of-view of the victim and passive network observers, there are no abnormal artifacts associated with the network connection that could lead to detection.
\item Dynamic linker hijacking and supply chain attacks will result in a computationally efficient method to rewrite HTTP requests, lowering the risk of abnormal memory or CPU spikes typically associated with decrypting and processing network traffic.
\end{enumerate}

While overriding DNS responses on the endpoint or directly exfiltrating the decrypted data to an attacker-owned server may accomplish similar goals to our attack, both methods would introduce artifacts that make identifying the attacker's actions possible.

\subsection{Attack}

Before AWS addressed the proposed attack, we verified that the end-to-end attack worked on AWS CloudFront domains of the form \texttt{[*].cloudfront.net} using Firefox, and it should also work with any hosting provider that supports domain fronting or wildcard certificate fronting, and any client application susceptible to execution flow hijacking or supply chain attacks. This section focuses on CloudFront, but we have also verified that the attack works against most Fastly domains, e.g., we were able to successfully intercept Firefox connections to \texttt{www.reddit.com}.

In our proof-of-concept illustrated in Figure \ref{fig:attack-poc}, the goal of the attacker is to create a stealthy command-and-control channel while also eavesdropping on all the victim's encrypted connections to a domain of the form \texttt{[*].cloudfront.net} in a way that is completely opaque to the user and network monitoring tools. The attack is straightforward:
\begin{enumerate}
\itemsep0em
\item The attacker creates an EC2 instance to act as the origin server and installs a webserver capable of proxying traffic, e.g., \texttt{nginx}. The origin server can run on any hosting provider, preferably closer to the CDN to reduce latency.
\item The attacker then creates a CloudFront distribution point, configures the above EC2 instance as the origin server, and configures CloudFront to forward all request headers to the origin server.
\item In the Fastly case, the attacker would then register a domain name of the same length as the domains to be intercepted and configure the CNAME record to point to Fastly.
\item On the victim's endpoint, the attacker uses dynamic linker hijacking to intercept popular function calls related to encryption, e.g., \texttt{PR\_Write} and \texttt{PR\_read}. We modified \texttt{mercury}'s \cite{mercury} intercept functionality for this step and provide a code sample in Appendix \ref{app:attack-source-code}.
\item During interception, if the HTTP \texttt{Host} contains a pattern consistent with default CloudFront distribution points, the attacker rewrites the \texttt{Host} to point to the attacker-owned distribution point.
\item In order to know the victim's original destination, the attacker identifies common HTTP headers such as the \texttt{User-Agent} field and rewrites the data associated with those headers to include the original destination. The \texttt{nginx} configuration code to extract this value is given in Appendix \ref{app:attack-source-code}.
\item The attacker optionally rewrites more data in the HTTP request/response headers to facilitate stealthy command-and-control.
\item When the request reaches the attacker's origin server, the attacker records all relevant information and proxies the request on behalf of the victim.
\end{enumerate}
When testing this attack, Firefox would segfault if we attempted to create a larger HTTP request by changing the length of the HTTP \texttt{Host} or creating additional HTTP headers. The proof-of-concept worked around this constraint by not modifying the length of the initial HTTP request. This constraint is easily overcome if the attacker registers multiple domains of varying length and replaces the HTTP \texttt{Host} with a domain of appropriate length. From the point-of-view of the victim and non-CDN network monitoring tools, all data features will be legitimate: the TLS \texttt{client\_hello} contains the victim's intended CloudFront domain, CloudFront returns a valid certificate, the IP address is the same, the response content exactly matches what the victim requested, and the browser/OS would not be able to log the malicious domain.

The CDN may have enough information to detect this attack, but we are unaware of any CDN that currently makes this data available to end users.

\section{Discussion}
\label{sec:discussion}

The scanning results presented in Section \ref{sec:measurement-results} were meant to be representative, but not necessarily exhaustive. For example, there are many AWS services that map to the \texttt{*.amazonaws.com.} canonical name, but we only gave specific results for Elastic Load Balancer. The methodology of Section \ref{sec:measurement} can be used for future studies that further characterize domain name misinformation and its support.

We have shown that many popular CDNs support domain fronting in Section \ref{sec:measurement-results}, and that malicious actors can abuse that support to man-in-the-middle a victim's traffic as described in Section \ref{sec:attacking-df}. Again, the attack is possible because the unwitting victim is presented a valid certificate by the CDN and the attacker-proxied traffic is what the victim expected. New and developing standards may achieve the same goals as domain fronting without exposing users to the risks of our attack. For example, the Encrypted Client Hello (ECH) \cite{ech} Internet draft results in an encrypted \texttt{server\_name}. When used in combination with DNS-over-HTTPS \cite{doh} and TLS 1.3 \cite{tls13}, passive network observers would only be able to extract the destination's IP address, which would only provide CDN-level information. Importantly, privacy enhancing technologies would not need to misrepresent the domain name in the \texttt{server\_name} extension, and CDNs could more comfortably discontinue support for domain fronting. The connections between an earlier version of ECH, encrypted SNI, and censorship circumvention were previously established \cite{chai2019importance,frolov2019censorship,hoang2020assessing}.


\subsection{Ethics}

The study of domain name misinformation is naturally divisive because these techniques obfuscate key data features used by incident response teams to identify malware infections, while at the same time furthering privacy enhancing technologies. It is our hope that these results motivate the security and privacy research community to develop privacy enhancing technologies that are less prone to abuse by malicious actors.

While the attack in Section \ref{sec:attacking-df} may not be considered a traditional vulnerability, we do believe that hosting providers should be made aware that their support for domain fronting can facilitate such an attack. To further that goal, we sent a preprint of this paper to each named company in December 2022. Most companies responded within two months. We were able to provide additional data around the attack that helped AWS confirm and fix the root cause in CloudFront. Companies such as Vercel and Fastly acknowledged the attack and said that they are in the process of implementing additional controls.

\section{Conclusion}
\label{sec:conclusion}

Domain fronting, domain faking, and domainless fronting are domain name misinformation techniques, which all have serious implications for security and privacy due to the reliance on domain names in common security architectures and the importance of these techniques in privacy enhancing technologies. We have presented a novel measurement methodology to identify support among cloud infrastructure providers and shown that many content delivery networks and cloud infrastructure providers support domain name misinformation techniques; sometimes intentionally, sometimes optionally, and sometimes unwittingly.

We have also presented a straightforward attack that leverages dynamic linker hijacking and domain fronting in a way that would allow malicious actors to stealthily maintain a surveillance state. With our attack, the attacker is able to man-in-the-middle all the victim's encrypted traffic bound to a content delivery network that supports domain fronting, breaking the authenticity, confidentiality, and integrity guarantees expected by the user when using HTTPS. We have successfully demonstrated a working attack on most Fastly domains and \texttt{[*].cloudfront.net} domains, and the attack should apply to all domains hosted by a provider that supports domain fronting.


\bibliographystyle{plain}
\bibliography{domain_fronting_conf}


\appendices


\begin{center}
\begin{figure*}[t!]
\begin{mdframed}
\begin{Ccode}
PRInt32 PR_Write(PRFileDesc *fd, const void *buf, PRInt32 amount) {
    int native_fd = PR_FileDesc2NativeHandle(fd);
    if (fd_is_socket(native_fd)) {
        // stealthy domain front
        char *buffer_cast = (char *)buf;
        std::string fronted_buffer(buffer_cast);
        // set up regex and our CloudFront domain name
        std::string cloudfront_domain(".cloudfront.net");
        std::string attack_domain("d-------------.cloudfront.net");
        std::string user_agent_str("Mozilla/5.0 (X11; Ubuntu; Lin");
        std::regex cloudfront_regex("Host: ([a-z0-9]{14})[.]cloudfront[.]net");
        std::smatch match;
        if (std::regex_search(fronted_buffer, match, cloudfront_regex)) {
            std::string original_domain = match.str(1) + cloudfront_domain;
            if (fronted_buffer.find(original_domain) != std::string::npos) {
                fronted_buffer.replace(fronted_buffer.find(original_domain),
                                       original_domain.length(),
                                       attack_domain);
                fronted_buffer.replace(fronted_buffer.find(user_agent_str),
                                       user_agent_str.length(),
                                       original_domain);
                const void *fronted_buffer_c_str = fronted_buffer.c_str();
                invoke_original(PR_Write, fd, fronted_buffer_c_str, amount);
            }
        }
    }
    invoke_original(PR_Write, fd, buf, amount);                                                                                                          
} 
\end{Ccode}
\end{mdframed}
\caption{Interception code based on \texttt{mercury}'s \cite{mercury} intercept functionality to identify \texttt{*.cloudfront.net} domains and rewrite the HTTP request accordingly.}
\label{fig:prwrite}
\end{figure*}
\end{center}

\begin{center}
\begin{figure}
\begin{mdframed}
\begin{Lcode}
map $http_user_agent $new_host {
  default                   "";
  "~^(?<victim_host>.{14})" $victim_host;
}

location / {
  resolver 8.8.8.8;
  proxy_pass https://$new_host$request_uri;
  proxy_set_header Host $new_host;
}
\end{Lcode}
\end{mdframed}
\caption{\texttt{nginx} configuration to extract the victim's intended domain from the \texttt{User-Agent} field and proxy the victim's original connection.}
\label{fig:nginx-config}
\end{figure}
\end{center}

\section{Attack Source Code}
\label{app:attack-source-code}

The interception source code for the attack presented in Section \ref{sec:attacking-df} is based on \texttt{intercept.cc} in \texttt{mercury}'s GitHub's repository \cite{mercury}. Figure \ref{fig:prwrite} presents the modifications we made to the Netscape Portable Runtime's \cite{nspr} \texttt{PR\_Write} intercept function. A regular expression is used to match all \texttt{Host} headers containing a CloudFront domain of a specific length, which is then overwritten with the attacker controlled CloudFront domain. The victim's intended domain is written into the \texttt{User-Agent} string before invoking the original \texttt{PR\_Write} function.

The \texttt{nginx} configuration to extract the victim's intended domain from the \texttt{User-Agent} string and proxy the victim's traffic is given in Figure \ref{fig:nginx-config}.

\end{document}